\documentclass[11pt,a4paper]{amsart}

\usepackage{amsmath,amssymb,graphicx}
\usepackage{bm}

\newtheorem{definition}{Definition}
\newtheorem{defi}[definition]{Definition}
\newtheorem{lem}[definition]{Lemma}
\newtheorem{prop}[definition]{Proposition}
\newtheorem{thm}[definition]{Theorem}
\newtheorem{cor}[definition]{Corollary}
\newtheorem{ex}[definition]{Example}

\newcommand{\F}{\mathbb F}
\newcommand{\wt}{\mathrm{wt}_\lambda}
\newcommand{\wtBR}{\mathrm{wt}_{BR}}

\def\C{\mathbb{C}}
\def\Cc{\mathcal{C}}

\def\Z{\mathbb{Z}}
\def\tr{\mathop{\rm{tr}}\nolimits}

\def\bra#1{\langle #1|}
\def\braket#1#2{\langle #1|#2\rangle}
\def\ket#1{|#1\rangle}

\advance\marginparwidth5mm

\title[Subfield Metric and Application to Quantum Error Correction]{The Subfield Metric and its Application to Quantum Error Correction}

\begin{document}

\author{Markus Grassl}
\address{International Centre for Theory of Quantum Technologies,
  University of Gdansk,
  Poland}
  \email{markus.grassl@ug.edu.pl}

\author{Anna-Lena Horlemann}
\address{School and Institute of Computer Science,
  University of St.Gallen,
  Switzerland}
  \email{anna-lena.horlemann@unisg.ch}

\author{Violetta Weger}
\address{Institute for Communications Engineering,
  Technical University of Munich,
  Germany}
  \email{violetta.weger@tum.de}

\begin{abstract}
We introduce a new weight and corresponding metric over finite extension fields for asymmetric error correction. The weight distinguishes between elements from the base field and the ones outside of it, which is motivated by 
asymmetric quantum codes.
We set up the theoretic framework for this weight and metric, including upper and lower bounds, asymptotic behavior of random codes, and we show the existence of an optimal family of codes achieving the Singleton-type upper bound. 
\end{abstract}

\maketitle

\section{Introduction}

The results presented in this article are motivated by the problem to develop a framework that links asymmetric quantum error-correcting codes and classical coding theory.  

Asymmetric quantum codes have their origin in the observation that in many physical systems, different types of errors occur with different probabilities.  For a qubit system, the errors are bit flips ($X$-errors), phase flips ($Z$-errors), as well as their combination ($Y$-errors) (see e.g. \cite{surv}). Most asymmetric quantum codes in the literature are based on the so-called CSS construction \cite{CaSh96,Ste96:simple}, which allows to correct $X$- and $Z$-errors independently.  The number of errors of $X$-type and of $Z$-type grows linearly with the dimension of the individual quantum system (qudit), while the number of combinations grows quadratically. Treating $X$- and $Z$-errors independently introduces a third type of errors, namely their combinations. As discussed in more detail in Section \ref{sec:asymmetricQECC} below, it is more natural to distinguish only two types of errors: diagonal errors and  non-diagonal ones. This can be modelled using codes over a quadratic extension field $\F_{q^2}$ and distinguishing errors in the base field $\F_q$ or its set-complement $\F_{q^2}\setminus \F_q$.  We consider the more general setting of extension fields $\F_{q^m}$ with arbitrary $m>1$.

To cope with the problem introduced by quantum error-correction, we introduce a new metric, called the $\lambda$-subfield metric. This metric gives errors that live outside of the base field a larger weight than errors in the base field, which are considered more common in this scenario. This allows us to correct more errors from the base field than by simply considering the Hamming metric.  

The paper is structured as follows. In Section \ref{sec:prelim} we recall the basics of classical coding theory and in Section \ref{sec:asymmetricQECC} the required background of quantum error-correcting codes. In Section \ref{sec:subfield} we introduce the new $\lambda$-subfield metric, which will be the main object of this paper. We derive the classical bounds, such as Singleton-type bounds, Plotkin-type bounds and a Gilbert-Varshamov-type bound for the new metric in Section \ref{sec:bounds}.  In Section \ref{sec:MRD-lambda} we study the maximum $\lambda$-subfield distance codes and in Section \ref{sec:mac} we present the MacWilliams identities for the case $m=2$, which is of particular interest for quantum error-correcting codes. Finally, we conclude this paper in Section \ref{sec:concl}.

\section{Preliminaries}\label{sec:prelim}

Throughout the paper $q$ is a prime power and $\F_{q^m}$ denotes the finite field of order $q^m$. Any $\mathcal{C} \subseteq \mathbb{F}_{q^m}^n$ is a code over $\F_{q^m}$ of length $n$. We call $\mathcal{C} \subseteq \mathbb{F}_{q^m}^n$ an $[n,k]$ linear code, if it is a $k$-dimensional subspace of $ \mathbb{F}_{q^m}^n$.

Several metrics can be defined on $\F_{q^m}^n$, among which we will use the Hamming metric $d_H$, defined as
$$ d_H(x,y):= |\{i \in \{1, \ldots, n\} \mid x_i \neq y_i\}|,$$
and the rank metric $d_R$, defined as
$$ d_R(x,y) := \dim_{\F_q} \langle x_1-y_1, \dots, x_n-y_m \rangle,$$
where $\langle x_1-y_1, \dots, x_n-y_n \rangle$ is the $\F_{q}$-vector space generated by the $x_i-y_i$. 
The respective weights are defined as the distances to the origin, i.e.,
$$ wt_H (x):= d_H(x,0) \quad, \quad wt_R(x):= d_R(x,0) .$$

For a code $\Cc\subseteq \F_{q^m}^n$ the minimum Hamming (respectively rank) distance $d_H(\Cc)$ (respectively $d_R(\Cc)$)  is defined as the minimum of the pairwise distances of elements of the code. 

The minimum distance is of great importance in coding theory, as it directly indicates how many errors a code can correct.

The Singleton-type bounds give an upper bound on the minimum distance of a code and state that
$$ \log_{q^m}(|\Cc|) \leq n-d_H(\mathcal{C})+1 ,$$
$$ \log_{q}(|\Cc|) \leq \max(m,n)(\min(m,n)-d_R(\mathcal{C})+1 ).$$
Codes achieving these bounds are called MDS (maximum distance separable) codes in the Hamming metric, respectively MRD (maximum rank distance) codes in the rank metric. It is well-known that MDS codes exist if $q^m\geq n-1$ and MRD codes for any set of parameters, see e.g., \cite{ga85a,ma77}.

Apart from the Hamming and the rank metric also several other metrics have been introduced to coding theory, often considering a particular channel to cope with. In this paper, we will introduce a new metric, which is suitable for quantum error correction, which we cover in the next section.

\section{Asymmetric Quantum Codes}\label{sec:asymmetricQECC}
We now give some background on quantum error-correcting codes. For an overview, see for example \cite{Gra21,KKKS06}. At the end of this section we motivate the main idea of this paper, i.e., to distinguish errors from the base field from those outside of it, for possible applications in quantum error correction.

For the complex vector space $\C^q$ of dimension $q$, we
label the elements of an orthonormal basis by the elements of the
finite fields, i.\,e.,
\begin{alignat*}{5}
  \mathcal{B} =\{ \ket{\alpha}\colon\alpha\in \F_q\}.
\end{alignat*}
An orthonormal basis of the dual vector space $(\C^q)^*$ is denoted by
\begin{alignat*}{5}
  \mathcal{B}^* =\{ \bra{\beta}\colon\beta\in \F_q\},
\end{alignat*}
such that $\braket{\beta}{\alpha} = \delta_{\beta,\alpha}$.

On the space $\C^q$, we define the following operators
\begin{alignat*}{5}
                    X^\alpha &{}= \sum_{x\in\F_q} \ket{x+\alpha}\bra{x},\\
  \text{and}\qquad  Z^\beta &{}= \sum_{y\in\F_q} \omega^{\tr(\beta y)}\ket{y}\bra{y},
\end{alignat*}
where $\omega=\exp(2\pi i/p)$ is a complex primitive $p$th root of
unity and $\tr(y)$ denotes the absolute trace from $\F_q$ to $\F_p$.
The operator $X^\alpha$ corresponds to a classical additive error
mapping the basis state $\ket{x}$ to the basis state $\ket{x+\alpha}$.
It is referred to as \emph{generalized bit-flip error} or $X$-error.
The operator $Z^\beta$, which is referred to as \emph{phase error} or
$Z$-error, does not have a direct classic correspondence.  Note that
the operators $\{Z^\beta\colon\beta\in\F_q\}$ are diagonal, while the
diagonal of any operator $X^\alpha Z^\beta$ with $\alpha\ne 0$ is
zero.

The set of $q^2$ operators
\begin{alignat}{5}
  \mathcal{E} = \{ X^\alpha Z^\beta\colon
  \alpha,\beta\in\F_q\}\label{eq:errorbasis}
\end{alignat}
is an orthogonal basis of the vector space of operators on $\C^q$ with
respect to the Hilbert-Schmidt inner product. Hence, any linear
operator can be expressed as a linear combination of these error operators.

For the $n$-fold tensor product $(\C^q)^{\otimes n}=\C^q\otimes\ldots\otimes\C^q$,
we define the error operators on
\emph{$n$ qudits}
\begin{alignat}{5}
  X^{\bm{a}}Z^{\bm{b}} &=X^{a_1}Z^{b_1}\otimes\ldots\otimes X^{a_n}Z^{b_n},\label{eq:error_operators}
\end{alignat}
with $\bm{a}=(a_1,\ldots,a_n)\in\F_q^n$ and $\bm{b}=(b_1,\ldots,b_n)\in\F_q^n$.
The weight of an error is defined as the number of tensor factors that
are different from identity.
A quantum error-correcting code $\mathcal{Q}$ is a subspace of the complex Hilbert space $(\C^q)^{\otimes n}$. By $[\![n,k,d]\!]_q$ we denote such a code $\mathcal{Q}\le(\C^q)^{\otimes n}$ of dimension $\dim(\mathcal{Q})=q^k$, and refer to $n$ as its length. Similar as for classical codes, a quantum code of minimum distance $d$ can detect any error of weight at most $d-1$, or the error has no effect on the states in the quantum code.

The \emph{depolarizing channel} is the quantum analog of a discrete
uniform symmetric channel. It either transmits a quantum state
faithfully, or outputs a completely random (maximally mixed) quantum
state. In terms of the error operators \eqref{eq:errorbasis}, any
non-identity operator is applied with equal probability. In the
literature, there are also quantum codes which are designed for the
case when the probabilities are non-identical. Such a case of
\emph{asymmetric quantum codes} has first been discussed for the case
of qubits, i.e., $q=2$ \cite{IoMe07}.  The so-called CSS construction
\cite{CaSh96,Ste96:simple} results in quantum codes for which the
correction of $X$- and $Z$-errors can be performed independently.  CSS
codes can be designed to correct a different number of $X$-
and $Z$-errors.  Constructions of asymmetric CSS codes have been
considered for larger dimensions $q$ as well (see, e.g., \cite{EJLP13}).

The \emph{dephasing channel} is a quantum channel for which only
$Z$-errors occur with equal probability.  For many quantum systems,
such phase errors are more likely than general errors.  This can be
modelled as the combination of a dephasing channel with a depolarizing
channel.  While asymmetric CSS codes can be designed to correct more
$Z$- than $X$-errors, such codes distinguish three types of errors:
$Z$-errors, $X$-errors, and their combination. A more adequate model
is to distinguish between $Z$-errors and other errors, i.e., between diagonal
error operators and their linear complement.  In terms of the error
operators $X^\alpha Z^\beta$, we distinguish errors corresponding to
$(0,\beta)$, $\beta\ne 0$ from $(\alpha,\beta)$ with $\alpha\ne 0$.
Identifying $\F_q\times \F_q$ with $\F_{q^2}$ via
$(\alpha,\beta)\mapsto \varepsilon=\alpha\gamma+\beta$ for some fixed
$\gamma\in\F_{q^2}\setminus\F_q$, this results in a
distinction between errors in the base field $\F_q$ and those errors
that generate the extension field, i.e., $\varepsilon\in\F_{q^2}\setminus\F_q$.
This differs from the CSS approach where one distinguishes between $(0,\beta)$ with $\beta \neq 0$, i.e., $Z$-errors, and $(\alpha,0)$, with $\alpha \neq 0$, i.e., $X$-errors.

\section{The Subfield Weights and Metrics}\label{sec:subfield}

To distinguish the two types of possible errors (inside and outside the base field) we introduce the following two weights. In the first we give different weightings to the two types, whereas in the second weight we count the two types of errors separately. The first then gives rise to a proper metric on $\mathbb F_{q^m}^n$, while the second carries more detailed information about the structure of the codewords.

\subsection{The $\lambda$-subfield metric}

\begin{defi}
Let $\lambda\geq 1/2$. We define the \emph{$\lambda$-subfield weight} on $\F_{q^m}$ as
$$ \wt (x) :=
\begin{cases} 
  0 & \text{if $x=0$,} \\ 
  1 & \text{if $x\in \F_q\setminus \{0\}$,}\\ \lambda & \text{if $x\in \F_{q^m}\setminus\F_q$,} \end{cases}$$
and extend it additively coordinate-wise to $\F_{q^m}^n$.

The \emph{$\lambda$-subfield distance} between $x,y \in \F_{q^m}^n$ is defined as $$ d_\lambda(x,y):= \wt(x-y) .$$
\end{defi}

We remark that the $\lambda$-subfield weight partitions the ambient space $\F_{q^m}$ according to the different weights given to the elements. This is similar to other weights, like the homogeneous weight or the one induced by the Sharma-Kaushik metric (see e.g. \cite{gabmetrics}) and has been studied extensively in the literature (see e.g. \cite{GlLu15}). 
However, the specific case of partitioning the finite field into a subfield and the remaining elements has not received particular attention before. Furthermore, for $\lambda=1$ we recover the Hamming weight as a special case.

\begin{lem}\label{lem:lambda_metric}
The $\lambda$-subfield distance is a metric on $\F_{q^m}^n$. 
\end{lem}
\begin{proof}
We prove the case $n=1$. The other cases are implied by the additivity of the weight.

By definition we have $d_\lambda(x,y)=0$ if and only if $x=y$.
Moreover, it is clearly symmetric, since $x\in \F_q$ if and only if $ -x \in \F_q$.

Now let $x,y,z \in \F_{q^m}$. For the triangle inequality note the following:
$$ d_\lambda(x,z)= \begin{cases}
0 & \text{if $x=z$,} \\ 1 & \text{if $x-z\in \F_q\setminus \{0\}$,} \\ \lambda & \text{if $x-z\in \F_{q^m}\setminus \F_q$.} 
\end{cases} $$
\begin{itemize}
\item
For $x=z$ we hence clearly have that $d_\lambda(x,z)=0 \leq d_\lambda(x,y) + d_\lambda(y,z)$. 
\item
For $x-z \in \F_q\backslash \{0\}$ and $x,y, z$ all distinct, we have that $d_\lambda (x,z)=1$ and 
$d_\lambda(x,y) , d_\lambda(y,z) \geq \min(\lambda,1) \geq  1/2$, which implies 
$$ d_\lambda(x,y) + d_\lambda(y,z) \geq \frac{1}{2}+ \frac{1}{2} = 1 = d_\lambda (x,z) .$$
\item
For $x-z \in \F_{q^m}\backslash \F_q$ we have that at least one of
$x-y,z-y$ must be in $\F_{q^m}\backslash \F_q$, otherwise $x-z =
(x-y)-(z-y)$ could not be in $\F_{q^m}\backslash \F_q$. This again
implies that $$d_\lambda(x,z)= \lambda \leq d_\lambda(x,y) +
d_\lambda(y,z).$$
\end{itemize}
\end{proof}

We remark that the second point in the proof of the triangle inequality above would not generally be true if $\lambda < 1/2$, which explains the restriction for $\lambda$.

The minimum distance $d_\lambda(\mathcal C)$ of a code $\mathcal C$ is defined as usual, as the minimum of
the pairwise distances.  Then
we get the classical error-correction capability of a code as follows.

\begin{lem}
Let $\mathcal{C}\subseteq \F_{q^m}^n$ be a code with minimum $\lambda$-subfield distance $d$. Then any error vector of $\lambda$-weight at most $\lfloor \frac{d-1}{2}\rfloor$ can uniquely be corrected, i.e., there is one unique closest codeword $c\in\mathcal{C}$ to a received word $r=c+e$ if $\wt(e)\leq \lfloor \frac{d-1}{2}\rfloor$.
\end{lem}
\begin{proof}
Let $e\in \F_{q^m}^n$ be an error vector of $\lambda$-subfield weight at most $\lfloor \frac{d-1}{2}\rfloor$. Assume by contradiction that there are two codewords $c,c'\in \mathcal C$ with $d_\lambda (c,c+e), d_\lambda(c',c+e) \leq \lfloor \frac{d-1}{2}\rfloor$. Then we get, by the triangle inequality,
$$ d_\lambda(c,c')\leq d_\lambda (c,c+e)+d_\lambda (c',c+e) \leq d-1,$$
which contradicts the fact that $d_\lambda(\mathcal C)=d$.
\end{proof}

Note that the $\lambda$-subfield weight of a vector $x \in \mathbb{F}_{q^m}^n$ does generally not prescribe the Hamming weight, nor how many entries in $x$ are from the base field and how many are from the extension field. To capture exactly this information we define the base-roof subfield weight in the following.

\subsection{The base-roof (BR)-subfield weight}

Since the $\lambda$-subfield weight is completely determined by the number of entries in the base field and the number of entries which lie exclusively in the extension field, that is, not in the base field, we will introduce two functions related to these two numbers. Thus, we partition the ambient space and weight the blocks accordingly. this is a well-known technique, e.g. \cite{GlLu15}.

Due to their nature, we call an entry $x_i$ of a  vector $x \in
\F_{q^m}^n$ of \emph{base type} if it is an element of the base field
$\F_q$, respectively of \emph{roof type} if it is
an element of the extension field, but not in the base field.  
\begin{defi}
For $x \in \mathbb{F}_{q^m}^n$ we define its  \emph{base weight} to be 
$$\text{wt}_b(x):=|\{i \in \{1, \ldots, n\} \mid x_i \in \mathbb{F}_q \setminus \{0 \} \}| $$ 
and its \emph{roof weight} as
$$\text{wt}_r(x):= |\{ i \in \{1, \ldots, n\} \mid x_i \in \mathbb{F}_{q^m} \setminus \mathbb{F}_q\}|.$$ 
By abuse of notation, for $x,y \in \mathbb{F}_{q^m}^n$, we define the \emph{base distance} between $x$ and $y$ as
\begin{align*}
   d_b(x,y) &:=\text{wt}_b(x-y),
   \end{align*} 
and the \emph{roof distance} as
\begin{align*}
   d_r(x,y) &:= \text{wt}_r(x-y).
\end{align*} 
\end{defi}

Note that the above weight functions are not weights that induce  distances. In fact, for a weight function to induce a distance it necessarily needs to be positive definite, symmetric, and it has to satisfy the triangle inequality. 
For a vector $x
\in \mathbb{F}_{q^m}^n$, neither $\text{wt}_b(x)=0$ nor
$\text{wt}_r(x)=0$ implies that $x=0$. For example, $x$ could live in
$(\mathbb{F}_{q^m} \setminus \mathbb{F}_q)^n$ and have
$\text{wt}_b(x)=0$. Similarly, any $x \in \mathbb{F}_q^n$ has roof
weight $0$, without being the zero vector.

Therefore, the base and roof distances are not metrics on $\F_{q^m}^n$. 
However, we have the following properties of these `distances'.
\begin{lem}
Let $x,y,z \in \mathbb{F}_{q^m}^n$. Then,
\begin{enumerate}
    \item $d_b(x,y)=0$ if and only if $x=y$ or $x-y \in (\mathbb{F}_{q^m} \setminus \mathbb{F}_q)^n$ and $d_b(x,y) \geq 0 $ for all $x,y \in \mathbb{F}_{q^m}^n.$
    \item $d_b(x,y)=d_b(y,x)$  for all $x,y \in \mathbb{F}_{q^m}^n.$
    \item $d_r(x,y) = 0$ if and only if $x-y \in \mathbb{F}_q^n$ and $d_r(x,y) \geq 0$  for all $x,y \in \mathbb{F}_{q^m}^n.$
    \item $d_r(x,y) = d_r(y,x)$  for all $x,y \in \mathbb{F}_{q^m}^n.$
    \item $d_r(x,y) \leq d_r(x,z) + d_r(z,y)$  for all $x,y,z \in \mathbb{F}_{q^m}^n.$
\end{enumerate}
The roof distance is hence a pseudometric on $\F_{q^m}^n$.
\end{lem}
\begin{proof}
The first four points are straightforward. We only have to prove that the roof distance satisfies the triangle inequality. For this note that it is enough to consider $n=1$ due to the additivity of the distance. Since the roof distance is induced by the roof weight, it is enough to prove the triangle inequality for the roof weight, that is  for $x,y \in \mathbb{F}_{q^m}$ we have $$  \text{wt}_r(x-y+z-z) \leq \text{wt}_r(x-z)+\text{wt}_r(z-y),$$ since then 
\begin{align*} d_r(x,y) = \text{wt}_r(x-y)  & = \text{wt}_r(x-y+z-z) \leq \text{wt}_r(x-z)+\text{wt}_r(z-y) \\ & = d_r(x,z) +d_r(z,y). \end{align*}
If $x-y \in \mathbb{F}_q$ we get the inequality trivially as $\text{wt}_r(x-y)=0.$ Thus, we can assume that $x-y \in \mathbb{F}_{q^m} \setminus \mathbb{F}_q$. This implies that  $x$ or $y$ live in $ \mathbb{F}_{q^m}\setminus \mathbb{F}_q$, hence the inequality is trivially satisfied as well. 
\end{proof}

Note that the triangle inequality does not hold for the base distance. 
\begin{ex}
  Let us consider $\mathbb{F}_2[\alpha]$ with $\alpha^2=\alpha+1.$ Let $x=\alpha, y= \alpha+1$ and $z=0.$ Then
  $$d_b(x,y)= \text{wt}_b(x-y) = \text{wt}_b(1) = 1,$$ but 
  $$d_b(x,z) + d_b(z,y)= 0.$$
\end{ex}

\begin{defi}
We define the  \emph{base-roof (BR)-weight} of $x$ as
$$ \mathrm{wt}_{BR}(x) :=(\text{wt}_b(x), \text{wt}_r(x)) .$$
Analogously, we define the  \emph{BR-distance} as
$$ d_{BR}(x,y) :=( d_b(x,y), d_r(x,y)) , $$
 for $x,y \in \mathbb{F}_{q^m}^n$.
\end{defi}

The BR-weight thus acts as the \emph{composition vector}, which counts the number of coordinates in each block. Hence, the BR-weight
completely determines the $\lambda$-subfield weight. That is, if $\text{wt}_{BR}(x) = (s,t)$, then  $\wt(x)=s+\lambda t.$
Note that the BR-distance is not
an actual distance (in particular, since its codomain is $\mathbb
N_0^2$),
however, we do get the following properties, analogous to those of a distance.

\begin{prop}
Consider the following partial order on $\mathbb N_0^2$:
$$ (s,t) \preceq (s', t') :\iff s \leq s' \text{ and } t\leq t' .$$
Let $x,y\in \F_{q^m}^n$. Then 
\begin{enumerate}
    \item $d_{BR}(x,y)= (0,0)$ if and only if  $x=y$.
    \item $d_{BR}(x,y)\succ(0,0)$ for $x\neq y$.
    \item $d_{BR}(x,y)=d_{BR}(y,x)$.
    \item $d_{BR}(x,y) \not\succ d_{BR}(x,z) + d_{BR}(z,y)$ for $x, y, z \in \F_{q^m}^n$.
\end{enumerate}
\end{prop}
\begin{proof}
The first three properties easily follow from the definition of the distance and the partial order. It remains to show the variant of the triangle inequality.
Assume by contradiction that   $d_{BR}(x,y) \succ d_{BR}(x,z) + d_{BR}(z,y) .$ Then $d_b(x,y) \geq d_b(x,z)+d_b(z,y)$ and $d_r(x,y) \geq d_r(x,z)+d_r(z,y)$ (and at least one of them is a strict inequality). Then we get 
\begin{align*} d_\lambda(x,y) & = d_b(x,y) + \lambda d_r(x,y) \\ 
& > d_b(x,z)+d_b(z,y) + \lambda(d_r(x,z)+d_r(z,y)) \\ 
& = d_\lambda(x,z) + d_\lambda(z,y).
\end{align*}
  This is a contradiction to Lemma \ref{lem:lambda_metric}.
\end{proof}

We now adjust the definition of minimum distance of a code 
to this setting.

\begin{defi}\label{def:BRdistance}
    Let $\mathcal{C}\subseteq \F_{q^m}^n$ be a code. We define the
    \emph{set of BR-minimal distances} of the code to be the set
    \begin{align*} d_{BR}(\mathcal{C}) & :=  \min\{d_{BR}(x,y)\in \mathbb{N}_0^2 \mid x, y \in \mathcal{C}, x\neq y\}, \end{align*}
    where 
$\min(D):=\{(s,t)\in D\mid\nexists(s',t')\in D\colon (s',t')\prec(s,t)\}$.
In general, this will not be one unique pair of values, but several minima. 
\end{defi}

The minimum $\lambda$-subfield distance $d_\lambda (\mathcal C)$ of a code $C$ is determined by the minimal BR-distances $d_{BR}(\mathcal C)$ through
$$ d_\lambda(\mathcal C) = \min \{s+\lambda t \mid (s,t)\in d_{BR}(\mathcal C)\}.$$

However, note that the BR-weight of a vector does not uniquely determine the $\lambda$-subfield weight of a vector $x$. In particular, if $\wt(x)=u$, then the BR-weight of $x$ could be any $(s,t)$ such that $u=s+\lambda t$. This shows that the BR-weight carries more information about the codewords than the $\lambda$-subfield weight.

Even more, if we would use the BR-distance for error correction, any error that can be corrected with the minimum BR-distances of a code could also be corrected using the $\lambda$-subfield distance. Thus, when considering error correction, we will mostly focus on the minimum $\lambda$-subfield distance of a code.

\subsection{Examples of error correction in the subfield metric(s)}

We now present some examples of codes, their minimum distances in the different metrics, and their error correction and detection capability.

\begin{ex}
Consider $\F_4=\F_2[\alpha] $ with $\alpha^2=\alpha +1$, and the code $\mathcal C\subseteq \F_4^6$ generated by
$$ G= \begin{pmatrix} 1&1&1&\alpha&\alpha&\alpha \end{pmatrix} .$$
As the Hamming distance of this code is $6$, we could correct any $2$ errors in the Hamming metric. 
The codewords and their BR-weights are as follows:
\begin{align*}
\arraycolsep2\arraycolsep
\begin{array}{c|c}
c & \wtBR(c)\\
\hline
\begin{pmatrix} 0&0&0&0&0&0 \end{pmatrix} & (0,0) \\
\begin{pmatrix} 1&1&1&\alpha&\alpha&\alpha \end{pmatrix} & (3,3) \\
\begin{pmatrix} \alpha^2&\alpha^2&\alpha^2&1&1&1 \end{pmatrix}  & (3,3) \\
\begin{pmatrix} \alpha&\alpha&\alpha&\alpha^2&\alpha^2&\alpha^2 \end{pmatrix} & (0,6)
\end{array}
\end{align*}
We hence get 
$$ d_{BR}(\mathcal{C})  = \{(3,3), (0,6)\} .$$
In the $\lambda$-subfield metric, we get the minimum distance
$$d_\lambda(\mathcal{C})= \min\{ 3+3\lambda, 6\lambda\}.$$
For $\lambda \geq 1$, we get $d_\lambda=3+3\lambda$, which means we can correct any $b$ errors in $\mathbb{F}_q^*$ and any $r$ errors in $\mathbb{F}_{q^m}\setminus \mathbb{F}_q$, which are such that $b+\lambda r < \frac{3+3\lambda}{2}.$ 
For example, for $\lambda=2$ and $d_\lambda(\mathcal{C})=9$, we can correct \begin{itemize}
    \item $4$ base errors and $0$ roof errors,
    \item $2$ base errors and $1$ roof error, or
    \item $0$ base errors and $2$ roof errors.
\end{itemize}
\end{ex}
The above example shows, 
that the $\lambda$-subfield distance allows in particular more base errors to be corrected, compared to the Hamming metric.

\begin{ex}
    Consider $\F_4=\F_2[\alpha] $ with $\alpha^2=\alpha +1$, and the code generated by 
    $$G=\begin{pmatrix}
        1&0&\alpha\\
        0&1&\alpha^2
    \end{pmatrix} .$$
 The minimal BR-distances are 
    $$d_{BR}(\mathcal{C})  = \{(3,0), (1,1), (0,2)\}$$
    and the minimum $\lambda$-subfield distance is $$d_\lambda(\mathcal{C})=\min\{3,1+\lambda,  2\lambda\} , $$ i.e., for $\lambda \geq 2$ we get $d_\lambda(C)=3$. Thus, we can correct any error vector with one entry from the base field $\F_q^*$.

    Note that the Hamming distance of $\mathcal{C}$ is two, hence we could not correct any error with respect to the Hamming metric. E.g., the received word $(0,1,1)$ is Hamming distance one away from $(0,1,\alpha^2)$ and $(1,1,1)$, however in the $\lambda$-subfield distance its unique closest codeword is $(1,1,1)$.   
    
\end{ex}

\begin{ex}\label{ex:MDS16}
  Consider $\F_{16}=\F_2[\alpha]$ with $\alpha^4+\alpha+1=0$. The
  cyclic code $\mathcal{C}$ of length $17$ with generator polynomial $g(x)=x^4 +
  \alpha^{12}x^3 + \alpha^2x^2 + \alpha^{12}x + 1$ is an MDS code
  of length $n=17$, dimension $k=13$ and minimum Hamming distance $5$, i.e., we could correct any two errors over $\F_{16}$.  Its restriction to $\F_4$ has the same length, dimension $9$ and minimum Hamming distance $7$.

  The minimal BR-distances of this code are
  \begin{align*}
    d_{BR}(\mathcal{C})=\{(0,5), (1,4), (2,3), (3,2), (4,1), (7,0) \},
  \end{align*}
  which implies
  \begin{align*}
    d_\lambda(\mathcal{C})=\min\{5\lambda,1+4\lambda,2+3\lambda,3+2\lambda,4+\lambda,7\}.
  \end{align*}

  For $\lambda=3$, we get $d_\lambda(\mathcal{C})=7$, and the code can correct
  any error of $\lambda$-weight strictly less than $7/2$. The set of
  correctable errors includes up to $3$ errors from the base field
  $\F_4$, and single errors from $\F_{16}\setminus \F_4$, but not the
  combination of an error of base type and an error of roof type on
  different positions.

  The code $\mathcal{C}$ can be used to construct a quantum MDS code
  with parameters $[\![17,9,5]\!]_4$. Considered as a symmetric quantum
  code, it can correct two arbitrary errors. Considered as an
  asymmetric code, it can correct up to $3$ phase errors or a single
  general error (but not the combination of a phase error and
  a non-phase error on different positions).

\end{ex}

\begin{ex}\label{ex:GF2_17}
  Consider $\F_{4}=\F_2(\alpha)$ with $\alpha^2+\alpha+1=0$. The
  cyclic code $\mathcal{C}$ of length $17$ with generator polynomial $g(x)=x^8 + \alpha x^7 + \alpha x^5 + \alpha x^4 + \alpha x^3 + \alpha x + 1$ has length $n=17$, dimension $k=9$ and minimum Hamming distance $7$, i.e., it can correct any $3$ errors from $\F_{4}$.  Its restriction to $\F_2$ is the repetition code with dimension one and minimum Hamming distance $17$.

  The minimal BR-distances of this code are
  \begin{align*}
    d_{BR}(\mathcal{C})=\{  (0,7), (1,6), (2,5), (3,4), (4,3), (5,2), (8,1), (17,0)\},
  \end{align*}
  which implies
  \begin{align*}
    d_\lambda(\mathcal{C})=\min\{7\lambda, 1+6\lambda, 2+5\lambda, 3+4\lambda, 4+3\lambda, 5+2\lambda, 8+\lambda, 17\}.
  \end{align*}

  For $\lambda=2$, we get $d_\lambda=9$, and the code can correct $2$ roof errors, the combination of $1$ roof error with $1$ base error, as well as $4$ base errors.
  For $\lambda=3$, we get $d_\lambda=11$, and the code can correct the combination of $1$ roof error with $2$ base errors, as well as $5$ base errors.

  The code $\mathcal{C}$ can be used to construct a quantum code
  with parameters $[\![17,1,7]\!]_2$. Considered as a symmetric quantum
  code, it can correct three arbitrary errors.  Considered as an asymmetric code, it can, e.g., correct an arbitrary error combined with two phase errors, as well as five phase errors.
\end{ex}

\section{Upper and Lower Bounds in the $\lambda$-Subfield Metric}\label{sec:bounds}

In this section we will derive several upper and lower bounds for the codes in the $\lambda$-subfield metric. In particular, we will derive sphere packing and sphere covering bounds, Singleton-type and Plotkin-type bounds. Moreover, we show that random codes achieve some of them with high probability.

We will denote the size of the largest code in $\F_{q^m}^n$ with minimum $\lambda$-subfield distance $d$ by 
$$A_{q^m,\lambda}(n,d).$$
Since we will need them in the bounds, we first derive results about the volume of the balls in the $\lambda$-subfield metric, i.e., the number of vectors of a given $\lambda$-subfield weight.

\subsection{The volume of the balls in the $\lambda$-subfield metric}

We denote the spheres and balls around the origin of radius $r$ with respect to the $\lambda$-subfield distance by
\begin{align*} S_{r,\lambda}(\F_{q^m}^n) &:= \{ x \in \F_{q^m}^n \mid \wt(x) = r\} , \\
B_{r,\lambda}(\F_{q^m}^n) &:= \bigcup_{i=0}^r  S_{i,\lambda}(\F_{q^m}^n) , \end{align*}
respectively.

\begin{lem}\label{lemma:ballsize} Let $0 \leq r \leq \lambda n$  
be a positive integer. We have that
\begin{align*}
| B_{r,\lambda}(\F_{q^m}^n)| &= \sum_{j=0}^r \sum_{i=0}^{\lfloor \frac{j}{\lambda} \rfloor} (q^m-q)^i \binom{n}{i} (q-1)^{j-\lambda i} \binom{n-i}{j-\lambda i}. 
\end{align*}
\end{lem}
\begin{proof}
We first determine $ |S_{d,\lambda}(\F_{q^m}^n)| $. If $v\in \F_{q^m}^n$ with $\wt(v)=d$ has $i\leq d/\lambda$ entries of roof type, that is from $\F_{q^m}\backslash \F_q$, then it must have $d-\lambda i$ entries of base type, i.e., from $\F_q\backslash\{0\}$. 
Hence, there are
$$
(q^m-q)^i \binom{n}{i} (q-1)^{d-\lambda i} \binom{n-i}{d-\lambda i} 
$$
vectors in $\F_{q^m}^n$ with $i$ entries from $\F_{q^m}\backslash \F_q$, such that $\wt(v)=d$. Summing over all possible values for $i$, we get 
$$ |S_{d,\lambda}(\F_{q^m}^n)| =\sum_{i=0}^{\left\lfloor \frac{d}{\lambda} \right\rfloor} (q^m-q)^i \binom{n}{i} (q-1)^{d-\lambda i} \binom{n-i}{d-\lambda i} $$
which implies the statement by $|B_{r,\lambda}(\F_{q^m}^n)| = \sum_{j=0}^r  |S_{j,\lambda}(\F_{q^m}^n) |$.
\end{proof}

We will often be interested in the asymptotic size of the balls. For that we need to compute $\lim_{n\to \infty} \frac{1}{n} \log_{q^m}(|B_{r,\lambda}(\F_{q^m}^n)|).$

To compute the asymptotic size of the $\lambda$-subfield metric balls we will use the saddle point technique used in \cite{saddle}.
For this we consider two functions $f(x)$ and $g(x)$, both not depending on $n$, and define a generating function
\[ \Phi(x) = f(x)^n g(x). \]
For some positive integer $t$ we denote the coefficient of $x^t$ in $\Phi(x)$ by \[[x^t]\Phi(x).\] 

\begin{lem}[\text{\cite[Corollary 1]{saddle}}]\label{lemGardySole}
Let $\Phi (x) = f(x)^n g(x)$ with $f(0)\neq 0$, and $t(n)$ be a function in $n$. Set $T := \lim_{n\rightarrow \infty}t(n)/n$ and set $\rho$ to be the solution to 
\[ \Delta(x):= \frac{x f'(x)}{f(x)} = T .\]
If $ \Delta'(\rho) >0$, and the modulus of any singularity of $g(x)$ is larger than $\rho$, then for large $n$
\[ \frac{1}{n} \log_{q^m}( [x^{t(n)}]\Phi(x)) \approx \log_{q^m}(f(\rho)) - T \log_{q^m}(\rho) + o(1) .\]
\end{lem}

With this technique we can determine the asymptotic size of the balls in the $\lambda$-subfield metric:
\begin{thm}   Let us consider the radius $u$ of a ball as a function in $n$, and take $U:= \lim_{n \to \infty} u(n)/n.$
Further, let $f(x)=1+(q-1)x+(q^m-q)x^\lambda$ and $\rho$ be the solution to 
$$U = \frac{x(q-1)+\lambda(q^m-q)x^\lambda}{1+x(q-1)+(q^m-q)x^\lambda}.$$ Then the  size of the ball fulfills
    $$\lim\limits_{n \to \infty} \frac{1}{n} \log_{q^m} \left(|B_{u(n),\lambda}| \right) = \log_{q^m}(f(\rho))-U\log_{q^m}(\rho).$$
\end{thm}
\begin{proof}
   For the $\lambda$-subfield weight the function 
$$f(x)=1+(q-1)x+(q^m-q)x^\lambda$$
represents the number of elements of $\F_{q^m}$ of weight $0,1$ and $\lambda$, respectively (as the coefficients of the monomials with the corresponding degree). 
The generating function
$$\Phi(x)=f(x)^n$$
then represents the size of the balls via
$$[x^{r}]\Phi(x) =|B_{r,\lambda}(\F_{q^m}^n)|.$$
We let $\rho$ be the solution to 
\begin{align}\label{eq:U-rho}
    \frac{x(q-1)+\lambda(q^m-q)x^\lambda}{1+x(q-1)+(q^m-q)x^\lambda} =U.
\end{align}
Since $[x^{U}]\Phi(x)=|B_{U,\lambda}(\F_{q^m}^n)|$, the statement follows from Lemma \ref{lemGardySole}.
\end{proof}

\subsection{Sphere packing and sphere covering bounds}

We will start with the most intuitive upper and lower bounds, the sphere packing and sphere covering bound, which directly follow from Lemma \ref{lemma:ballsize}.

\begin{thm}\label{thm:sphere_cov_pack}
We have the following upper sphere packing and lower sphere covering bound:
\begin{align*}  
A_{q^m,\lambda}(n,d) & \leq \frac{|\F_{q^m}^n|}{ |B_{\left\lfloor \frac{d-1}{2} \right\rfloor,\lambda}(\F_{q^m}^n)|} = \frac{q^{mn}}{\sum_{j=0}^{\lfloor \frac{d-1}{2} \rfloor} \sum_{i=0}^{\lfloor \frac{j}{\lambda} \rfloor} (q^m-q)^i \binom{n}{i} (q-1)^{j-\lambda i} \binom{n-i}{j-\lambda i} } \\ A_{q^m,\lambda}(n,d) & \geq \frac{|\F_{q^m}^n|}{ |B_{d-1,\lambda}(\F_{q^m}^n)|} = \frac{q^{mn}}{\sum_{j=0}^{d-1} \sum_{i=0}^{\lfloor \frac{j}{\lambda} \rfloor} (q^m-q)^i \binom{n}{i} (q-1)^{j-\lambda i} \binom{n-i}{j-\lambda i} }.
\end{align*}
\end{thm}

Under certain restrictions on the dimension of the (linear) code, we can also prove that a random linear code achieves the sphere covering (or Gilbert-Varshamov-type) bound with high probability, if the field size is large enough.

\begin{thm}\label{thm:probGV}
Let us denote by  $g(d)=\log_{q^m}(|B_{d,\lambda}(\F_{q^m}^n)|)$.
Let $\varepsilon >0$ and $\mathcal{C}\subseteq \F_{q^m}^n$ be a randomly chosen linear code of dimension $k= \lceil(1- g(d)/n - \varepsilon)n \rceil$. Then the probability that $\mathcal{C}$ achieves the sphere covering bound, i.e., that its minimum $\lambda$-subfield distance is $d$, is at least $1-q^{m(1-\varepsilon n)}$.
\end{thm}
\begin{proof}
Let $c\in \mathcal{C}$ be randomly chosen, i.e., a random linear combination of the rows of the (random) generator matrix $G$. Then this is a random (uniformly distributed) non-zero element from $\F_{q^m}^n$. The probability that $\wt(c) < d$ is hence
$$\frac{ |B_{d-1,\lambda}(\F_{q^m}^n)|-1}{q^{mn}-1} \leq \frac{ |B_{d,\lambda}(\F_{q^m}^n)|}{q^{mn}} =q^{mg(d)-mn}.$$
We apply the union bound over all non-zero codewords and get that the probability that $\mathcal{C}$ has minimum distance less than $d$ is upper bounded by 
$$ q^{mk} q^{mg(d)-mn} \leq q^{m((1- g(d)/n - \varepsilon)n +1)} q^{mg(d)-mn}= q^{m(1-\varepsilon n)} .$$
This implies that the probability of $\mathcal{C}$ having minimum distance at least $d$ is at least $1-q^{m(1-\varepsilon n)}$.
\end{proof}

Note that the asymptotic behavior with respect to the sphere packing bound for general additive weights was studied in \cite{ho22dense:article}, which implies the following for our setting: 
\begin{thm}\cite[see Remarks 3.3 and 3.9]{ho22dense:article}

Let $\ell,m,n$ be positive integers with $\ell \mid m$.
\begin{enumerate}
    \item The probability that a random nonlinear code in $\F_{q^m}^n$ gets arbitrarily close (from below) to the lower bound in Theorem \ref{thm:sphere_cov_pack} goes to zero for growing $n$ or growing~$q^m$. 
    \item The probability that a random $\F_{q^\ell}$-linear code in $\F_{q^m}^n$ gets arbitrarily close (from below) to the lower bound in Theorem \ref{thm:sphere_cov_pack} goes to one for growing $q$ or growing~$\ell$.
\end{enumerate}
\end{thm}

We remark that the second part above also follows directly from Theorem \ref{thm:probGV}, for $\ell=m$.
\subsection{Singleton-type bound}

For the Singleton-type bound we need to assume that $\lambda \geq 1$.

\begin{thm}\label{thm:Singleton1}
Let $\lambda \geq 1$ and $\mathcal{C} \subseteq \mathbb{F}_{q^m}^n$ be a code (not necessarily linear) with minimum $\lambda$-subfield distance $d_\lambda$. Then
$$\left\lfloor \frac{d_\lambda -1}{\lambda}\right\rfloor \leq n-\lceil \log_{q^m}(\mid \mathcal{C}\mid)\rceil.$$
Thus for linear codes of dimension $k$, we get 
\begin{equation*}
\left\lfloor \frac{d_\lambda -1}{\lambda}\right\rfloor \leq n-k.
\end{equation*}
\end{thm}
\begin{proof}
 We clearly have that
$ \text{wt}_H(x) \leq \text{wt}_\lambda(x) \leq \lambda \text{wt}_H(x).$
Since $d_H(\mathcal{C})$ is an integer, for any  $ \lambda \geq 1$, we have $d_H(\mathcal{C}) \geq \left\lceil \frac{d_\lambda(\mathcal{C})}{\lambda} \right\rceil $, whenever
$d_\lambda(\mathcal{C}) \leq \lambda d_H(\mathcal{C})$. This yields
\begin{equation*}
\left\lfloor \frac{d_\lambda(\mathcal{C})-1}{\lambda} \right\rfloor \leq d_H(\mathcal{C}) -1 \leq n-k.
\end{equation*}
\end{proof}

We refer to codes that achieve the Singleton bound with equality as 
\emph{maximum $\lambda$-subfield distance (M$\lambda$D) codes}.
This implies that any code $\mathcal{C} \subseteq \mathbb{F}_{q^m}^n$ of dimension $k$ is a maximum $\lambda$-subfield distance code if its minimum $\lambda$-subfield distance $d_\lambda$ is
$$d_\lambda= \lambda(n-k)+\alpha,$$ 
for some $1\leq \alpha < \lambda+1$, i.e., for integer $\lambda$ we get $\alpha \in \{1, \ldots, \lambda \}$.

Note that, for (relatively) large $\lambda$, the bound can only be achieved if $n-k$ coordinates are exclusively from the extension field.

\begin{cor}\label{cor:Singleton_lin}
Let $\lambda \geq 1$ and $\mathcal{C} \subseteq \mathbb{F}_{q^m}^n$ be a linear code of dimension $k$ with minimum $\lambda$-subfield distance  $d_\lambda$. Then 
$$d_\lambda(\mathcal{C}) \leq \lambda(n-k)+1.$$
\end{cor}
\begin{proof}
    This follows easily from the fact, that we have a generator matrix which we can bring into systematic form. That is, there is at least one codeword $c \in \mathcal{C}$, that has $k-1$ zero entries and one entry being $1$. 
\end{proof}

\subsection{Plotkin bound}
Let us define the average $\lambda$-subfield weight of a code $\mathcal{C} \subseteq \mathbb{F}_{q^m}^n$ to be 
$$ \overline{\text{wt}}_\lambda(\mathcal{C})= \frac{1}{|\mathcal{C}|} \sum_{c \in \mathcal{C}} \text{wt}_\lambda(c).$$
Note that the average $\lambda$-subfield weight on $\mathbb{F}_{q^m}$ is given by 
$$ D:= \frac{q-1+ \lambda(q^m-q)}{q^m}= \lambda - (\lambda-1)q^{1-m}-q^{-m}.$$
Combining this with the usual Plotkin argument, i.e., \begin{equation}\label{eq:plotkin}
    d_\lambda(\mathcal{C}) \leq \frac{|\mathcal{C}|}{|\mathcal{C}|-1} \overline{\text{wt}}_\lambda(\mathcal{C}).
\end{equation}
we obtain the Plotkin bound for linear codes.
\begin{thm}
Let $\mathcal{C} \subseteq \mathbb{F}_{q^m}^n$ be a linear code, then 
$$d_\lambda(\mathcal{C}) \leq \frac{|\mathcal{C}|}{|\mathcal{C}|-1} n(\lambda - (\lambda-1)q^{1-m}-q^{-m}). $$ 
\end{thm}

\begin{proof} The claim easily follows from Equation \eqref{eq:plotkin} and since the average weight of a linear code can be bounded by $$\overline{\text{wt}_\lambda}(\mathcal{C}) \leq nD.$$ 
\end{proof}

Note that the Plotkin bound above implies the following bound on the size of the code, provided that the denominator 
is positive:
\begin{alignat}{5}
|\mathcal{C}|\le\frac{d_\lambda(\mathcal{C})}{d_\lambda(\mathcal{C})-n(\lambda-(\lambda-1)q^{1-m}-q^{-m})}\label{eq:Plotkin_size}
\end{alignat}
For fixed distance $d_\lambda(\mathcal{C})$, this bound is only valid for relative small length $n$. But we can use this bound to derive an upper bound on the size of codes $\mathcal{C}$ of unbounded length, provided that inequality \eqref{eq:cond_nprime} below holds. The main idea is to decompose the code $\mathcal{C}$ into shorter codes to which can apply the bound \eqref{eq:Plotkin_size}.

\begin{cor}
Let $\mathcal{C}$ be a linear code over $\F_{q^m}$ of length $n>n'$ with minimum
$\lambda$-subfield distance $d_\lambda(\mathcal{C})$, where
  \begin{alignat}{5}
    \frac{d_\lambda(\mathcal{C})}{\lambda}\le \left\lceil\frac{q^m d_\lambda(\mathcal{C})}{\lambda q^m-(\lambda-1)q-1}\right\rceil-1=:n'.\label{eq:cond_nprime}
  \end{alignat}
  Then
  \begin{alignat*}{5}
    |\mathcal{C}|    \le\frac{d_\lambda(\mathcal{C})q^{-mn'}}{d_\lambda(\mathcal{C})-n'(\lambda-(\lambda-1)q^{1-m}-q^{-m})}q^{mn}.
  \end{alignat*}
\end{cor}
\begin{proof}
  For each prefix $x\in\F_{q^m}^{n-n'}$, define the code $\mathcal{C}_x$ of
  length $n'$ by
  \begin{alignat*}{5}
    \mathcal{C}_x=\{ (c_{n-n'+1},&c_{n-n'+2},\ldots, c_n)\in\F_{q^m}^{n'}\colon\nonumber\\
    &(x_1,\ldots,x_{n-n'},c_{n-n'1+1},c_{n-n'+2},\ldots, c_n)\in \mathcal{C}\}.
  \end{alignat*}
  Each $\mathcal{C}_x$ is a (possibly empty) code of length $n'$ and minimum
  distance $d_\lambda(\mathcal{C})$. These codes are cosets of linear
  codes, i.e., they have the same distance distribution as linear
  codes. The length $n'$ is chosen such that the denominator in
  \eqref{eq:Plotkin_size} is positive, and we can apply that bound on the size of
  $\mathcal{C}_x$.  For $\mathcal{C}$, we get
  \begin{alignat*}{5}
    |\mathcal{C}|=\sum_{x\in\F_{q^m}^{n-n'}} |\mathcal{C}_x|&
    \le q^{m(n-n')} \frac{d_\lambda(\mathcal{C})}{d_\lambda(\mathcal{C})-n'(\lambda-(\lambda-1)q^{1-m}-q^{-m})}.
  \end{alignat*}
\end{proof}
When inequality \eqref{eq:cond_nprime} holds, we obtain a bound on the size of the code that is proportional to $q^{mn}$ for all $n>n'$.

To retrieve the same bounds for non-linear codes we have to consider
$\sum\limits_{x,y \in \mathcal{C}} d_\lambda(x,y)$
instead of $\sum\limits_{c \in \mathcal{C}} \text{wt}_\lambda(c).$
\begin{thm}
Let $\mathcal{C} \subseteq \mathbb{F}_{q^m}^n$ be a code with $d_\lambda(\mathcal{C}) > n(\lambda - (\lambda-1)q^{1-m} - q^{1-2m})$. 
Then, 
$$|\mathcal{C}| \leq \frac{ d_\lambda(\mathcal{C})}{  d_\lambda(\mathcal{C})-  n(\lambda - (\lambda-1)q^{1-m} - q^{-m}) }.$$
\end{thm} 
\begin{proof}
We observe that 
\begin{equation}\label{eq:Plotkin2} \sum\limits_{x,y \in \mathcal{C}} d_\lambda(x,y) \geq |\mathcal{C}|\left( |\mathcal{C}|-1\right) d_\lambda(\mathcal{C}). 
\end{equation}
Thus, it is enough to upper bound $\sum\limits_{x,y \in \mathcal{C}} d_\lambda(x,y)$, which we will do in the following.

For this let $A$ be the $|\mathcal{C}|\times n$ matrix having all codewords of $\mathcal{C}$ as rows.
Further, let $\mathcal{R}= \mathbb{F}_{q^m}\setminus \mathbb{F}_q$ be the set of roof elements of size $|\mathcal{R}|=q^m-q$. 
We use the notation $m_\ell$ for the number of appearances of any element $\ell \in \mathbb{F}_{q^m}$ in a column. We consider the cosets $T_\alpha:=\alpha+\F_q$ and partition $\F_{q^m}$ into $\F_q\cup \bigcup_{\alpha\in S}T_\alpha$ for some $S\subseteq \mathcal{R}$ of cardinality $q^{m-1}-1$. Then each of these sets $T_\alpha$ has cardinality $q$ and since they form a partition, we have that $T_\alpha \neq T_\beta$ for each $\alpha \neq \beta \in S$. Note that the distance between any pair of distinct elements in $T_\alpha$ is $1$, whereas the distance between elements of two different sets, i.e., $T_\alpha \neq T_\beta$, is $\lambda$. We hence get
\begin{align*}
    \frac{1}{n}\sum\limits_{x,y \in \mathcal{C}}d_\lambda(x,y)
    \leq{} &  \lambda\sum_{\substack{\alpha, \beta \in S\cup\{0\}\\ \alpha \neq \beta}}\sum_{\ell\in T_\alpha}\sum_{\ell' \in T_\beta} m_\ell m_{\ell'} +1  \sum_{\alpha \in S\cup \{0\}}\sum_{\substack{\ell, \ell' \in T_\alpha\\ \ell\neq \ell'}} m_\ell m_{\ell'} \\
    ={} &  \lambda(\sum_{\ell,\ell' \in \mathbb{F}_{q^m}} m_\ell m_{\ell'}- \sum_{\alpha \in S\cup\{0\}}\sum_{ \ell,\ell' \in  T_\alpha} m_\ell m_{\ell'})  \\ & +1 (  \sum_{\alpha \in S\cup\{0\}}\sum_{ \ell,\ell' \in T_\alpha} m_\ell m_{\ell'} - \sum_{\ell \in \mathbb{F}_{q^m}} m_\ell^2 )\\
     ={} &  \lambda\sum_{\ell,\ell' \in \mathbb{F}_{q^m}} m_\ell m_{\ell'}-(\lambda-1) \sum_{\alpha \in S\cup\{0\}}\sum_{ \ell,\ell' \in T_\alpha} m_\ell m_{\ell'}- \sum_{\ell \in \mathbb{F}_{q^m}} m_\ell^2 \\
     \leq{} & \lambda |\mathcal C|^2-(\lambda-1)|\mathcal C|^2q^{1-m}-|\mathcal C|^2q^{-m},
\end{align*}
where the last inequality follows from the fact that $\sum_{\alpha \in S\cup\{0\}}\sum_{ \ell,\ell' \in T_\alpha} m_\ell m_{\ell'}$ and $\sum_{\ell \in \mathbb{F}_{q^m}} m_\ell^2 $ are minimized if the $m_\ell$ are equal for all $\ell$, i.e., $m_\ell=q^{-m}|\Cc|$, whereas $\lambda\sum_{\ell,\ell' \in \mathbb{F}_{q^m}} m_\ell m_{\ell'}$ is maximized if $m_\ell=|\Cc|$ for one $\ell$. 
\end{proof}

\subsection{Comparison of bounds}
\begin{figure}[ht!]
$$
\includegraphics[width=6.5cm]{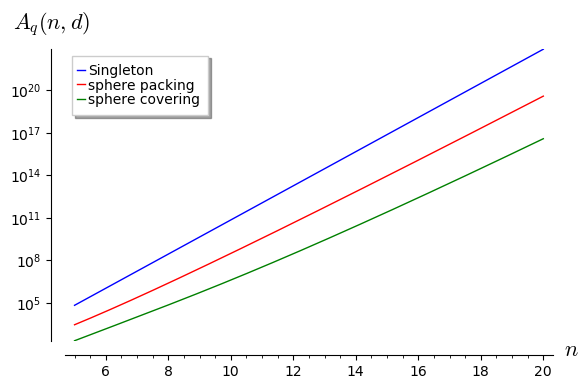}
\includegraphics[width=6.5cm]{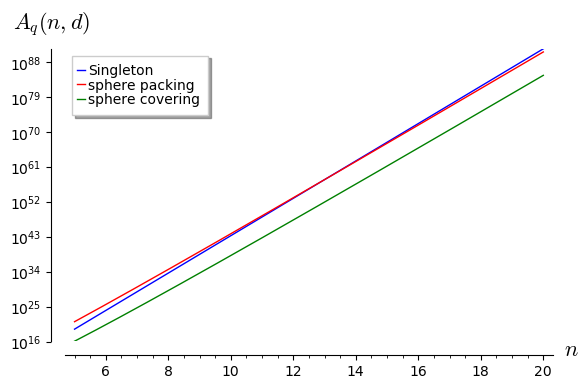}$$
\caption{Bounds on $A_{q^m,\lambda}(n,d)$ for  $q=4,m=2,\lambda=4,d=7$, and $q=4,m=8,\lambda=5,d=10$.\label{fig:plot1}}
\end{figure}

\begin{figure}[ht!]
$$\includegraphics[width=6.5cm]{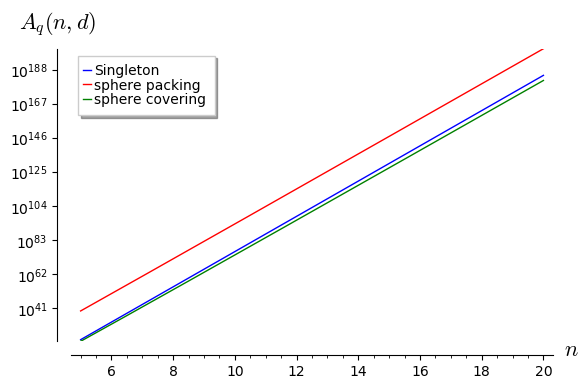}
\includegraphics[width=6.5cm]{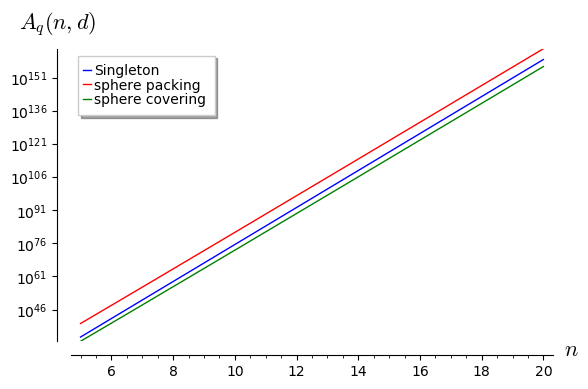}$$
\caption{Bounds on $A_{q^m,\lambda}(n,d)$ for  $q=521,m=4,\lambda=3,d=10$, and $q=5,m=12,\lambda=3,d=5$.\label{fig:plot2}}
\end{figure}

In Figures \ref{fig:plot1} and \ref{fig:plot2}, we plot the size $A_{q^m,\lambda}(n,d)$ of a code over $\F_{q^m}$ with $\lambda$-distance $d$ as a function of the code length $n$ for various choices of $q$, $m$, $\lambda$, and $d_\lambda$. 

As for the Hamming metric, none of the three bounds presented beats the others for all parameter sets. Generally, we see that for large field size (compared to the length) the Singleton bound is tighter than the others (which is supported by the fact that codes achieving this bound exist for large field extension degree, see Section~\ref{sec:MRD-lambda}). For smaller field size, however, we see that the sphere packing 
bound is tighter.

\begin{figure}[ht!]
$$\includegraphics[width=6.5cm]{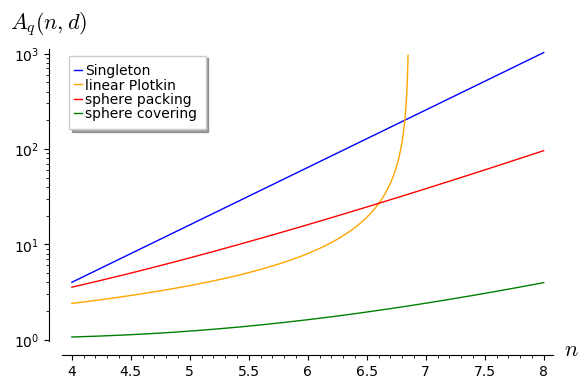}
\includegraphics[width=6.5cm]{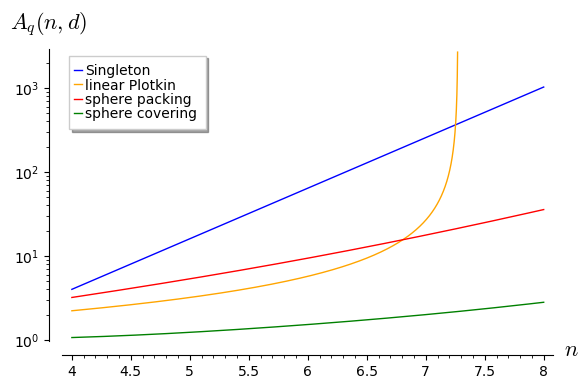}$$
\caption{Bounds on $A_{q^m,\lambda}(n,d)$ for  $q=521,m=4,\lambda=3,d=10$, and $q=5,m=12,\lambda=3,d=5$.\label{fig:plot3}}
\end{figure}

The Plotkin bound beats the other two upper bounds only for very short lengths of the code and large minimum distance, which we exemplify in Figure~\ref{fig:plot3}. Note that the Plotkin bound \eqref{eq:Plotkin_size} is only valid for relatively short codes.

\section{MRD Codes as M$\lambda$D Codes}\label{sec:MRD-lambda}

In this section we show that optimal codes in the rank metric are also optimal in the $\lambda$-subfield metric, in the sense that they achieve the Singleton-type bound from Theorem \ref{thm:Singleton1}. As for the Singleton-type bound itself, we assume $\lambda \geq 1$ in this section.

As we show in the following we can use the rank weight of a vector to lower bound its subfield weight.

\begin{lem}\label{lem:rank-subfield}
Let $v\in \F_{q^m}^n$ have rank weight $d$. Then $wt_\lambda(v)\geq \lambda (d-1) +1$.
\end{lem}
\begin{proof}
    If $wt_R(v)=d$, then at least $d-1$ entries in $v$ are in $\F_{q^m}\backslash \F_q$. These entries amount to a $\lambda$-subfield weight of $\lambda (d-1)$. Furthermore, there is at least one more non-zero entry, which adds weight $\lambda$ or $1$.
\end{proof}

\begin{thm}\label{thm:MRD-lambda}
Let $\mathcal{C}\subseteq \F_{q^m}^n$ be an MRD code. 
Then the minimal $\lambda$-subfield distance satisfies 
$$ \lambda (n-k) +1  \leq d_\lambda \leq \lambda(n-k+1)$$
for $\lambda\ge 1$.
Thus, the code achieves the Singleton-type bound and is an M$\lambda$D code.
\end{thm}
\begin{proof}
The upper bound on the distance follows from the Singleton-type bound. For the lower bound, assume by contradiction, that there was a non-zero codeword $x \in \mathcal{C}$ of $\wt(x) < \lambda(n-k)+1$. Then  $\mathrm{wt}_R(x) < n-k+1$, which is a contradiction to the MRD property. Thus, $d_\lambda(\mathcal{C}) \geq \lambda (n-k)+1$. 

We hence have 
$$   \lambda (n-k) +1 \leq d_\lambda \leq \lambda(n-k+1)
\iff \left\lfloor \frac{d_\lambda-1}{\lambda}\right\rfloor = n-k ,$$
which is exactly the Singleton-type bound with equality.
\end{proof}

Remember that Corollary \ref{cor:Singleton_lin} states a tighter Singleton-type bound for linear codes. This implies the following:

\begin{cor}
    Let $\mathcal{C}\subseteq \F_{q^m}^n$ be an MRD code. Then $d_\lambda(\Cc) = \lambda(n-k)+1$ and one of the minimal BR-distances of $\Cc$ is $(1,n-k)$.
\end{cor}

Therefore, an MRD code as an M$\lambda$D code has an error correction capability of at least $\frac{(n-k)\lambda}{2}$, i.e., for any $i\in \{0,1,\dots , \min\{\lfloor\frac{n}{\lambda}\rfloor, \lfloor \frac{n-k}{2} \rfloor\} \}$, we can correct
any $ \left\lfloor \frac{n-k}{2} \right\rfloor -i $
roof errors, plus
 any 
$ i\lambda$
base errors. 
Note that we require $i\lambda \leq n$, since there can never be more than $n$ errors of one type.
Theorem \ref{thm:MRD-lambda} implies that there are at least as many M$\lambda$D codes as MRD codes. Hence we can show that M$\lambda$D codes are dense, i.e., that the probability, that a randomly chosen code is an  M$\lambda$D codes, goes to one, whenever MRD codes are dense: 

\begin{cor}
Let $n$, $\ell, m$ be positive integers and consider $\F_{q^\ell}$-linear codes codes in $\F_{q^m}^n$ of minimum $\lambda$-subfield distance $2 \le d_\lambda \le \lambda n$. Then we have:
\begin{itemize}
\item The probability that a randomly chosen $\F_{q^\ell}$-linear code in $\F_{q^m}^n$ is an M$\lambda$D code
goes to one for growing $\ell$.
    
    \item If $m\geq n$ and $\ell  > (d-1)(n-d+1)$, then the probability that a randomly chosen $\F_{q^\ell}$-linear code in $\F_{q^m}^n$ is an M$\lambda$D code
goes to one for growing $q$.
    
    \item If $m < n$ and $\ell > (d-1)(n+m -d+1)-\ell \lceil n(d-1)/\ell \rceil$, then the probability that a randomly chosen $\F_{q^\ell}$-linear code in $\F_{q^m}^n$ is an M$\lambda$D code
goes to one for growing $q$.
\end{itemize}
\end{cor}
\begin{proof}
It was shown in \cite[Theorems 5.7 and 5.8]{ho22dense:article} that for all the parameters given in the statement MRD codes are dense. Together with Theorem \ref{thm:MRD-lambda} this implies the statement.
\end{proof}

\section{Subfield Weight Enumerators and MacWilliams Identity}\label{sec:mac}
In this section, we consider the special case of quadratic extensions
$\F_{q^2}$ which are related to the construction of quantum
error-correcting codes.

Recall from \eqref{eq:error_operators} that the error operators on
$n$ qudits are labelled by two vectors $\bm{a},\bm{b}\in\F_q^n$.  On
this space, we define the trace-symplectic form
\begin{alignat}{5}
  (\bm{a},\bm{b})*(\bm{a}',\bm{b}')=\sum_{i=1}^n \tr(a_ib_i'-a'_ib_i).\label{eq:symplectic_form}
\end{alignat}
For fixed $(\bm{a}',\bm{b}')$, this defines a character
\begin{alignat*}{5}
  \Psi_{(\bm{a}',\bm{b}')}\bigl((\bm{a},\bm{b})\bigr)=\omega^{(\bm{a},\bm{b})*(\bm{a}',\bm{b}')}
\end{alignat*}
on $\F_q^{2n}$ considered as an additive group (using notation from
\cite{ZiEr09}). Note, however, that in our case
$\Psi_x(y)=\Psi_y(x)^*$ (where ${}^*$ denotes complex conjugation) as
the trace-symplectic form is anti-symmetric, while Eq. (4) of
\cite{ZiEr09} uses a symmetric bilinear form.  Following the approach
of Delsarte \cite{Del72}, this setting can be used to prove
MacWilliams identities for the Hamming weight enumerators of an
additive code $\mathcal{C}$ and its dual $\mathcal{C}^*$ with respect
to the trace-symplectic form \eqref{eq:symplectic_form}.

Here we consider a refined weight enumerator based on the
partition
\begin{alignat}{5}
  P_0 &{}= \{ (0,0)\}\label{eq:partition0}\\
  P_1 &{}= \{ (0,b)\colon b\in\F_q\setminus\{0\}\}\\
  P_2 &{}= \{ (a,b)\colon a\in\F_q\setminus\{0\},b\in\F_q\}\label{eq:partition2}
\end{alignat}
of the alphabet $\F_q\times\F_q$. In terms of the error basis
\eqref{eq:errorbasis}, $P_0$ corresponds to the identity matrix, $P_1$
to the diagonal matrices excluding identity, and $P_2$ to the
non-diagonal matrices. Identifying $\F_q\times\F_q$ with $\F_{q^2}$,
$P_1$ corresponds to $\F_q\setminus\{0\}$ and $P_1$ to
$\F_{q^2}\setminus\F_q$. To devise the MacWilliams identities, we have to show that the partition $(P_0,P_1,P_2)$ is a so-called $F$-partition (Fourier-reflexive in the terminology of \cite[Theorem 3.5]{GlLu15}). By the criteria of the Lemma on p.~223 in \cite{ZiEr09}, this is equivalent to the conditions that $\sum_{x\in P_i} \Psi_x(y)$ only depends on the index $j$ of the
partition $P_j$ containing $y$, as well as that $\sum_{y\in P_j}
\Psi_x(y)$ only depends on the index $i$ of the partition $(P_0,P_1,P_2)$. 
In the following, $y=(a,b)\in\F_q\times\F_q$.
\begin{itemize}
  \item $i=0$: $\Psi_{(0,0)}(y)$ is the trivial character, and hence
    \begin{alignat*}{5}
      \sum_{x\in P_0} \Psi_x(y)=\Psi_{(0,0)}(y)=1.
    \end{alignat*}
  \item $i=1$: $\Psi_{(0,b')}(a,b)$ with $b'\ne 0$ depends only on $a$
    and hence defines a non-trivial character of $\F_q$. Then
    \begin{alignat*}{5}
      \sum_{x\in P_1}\Psi_x(y)=\sum_{b'\in\F_q\setminus\{0\}}\Psi_{(0,b')}(y)
      =\sum_{b'\in\F_q\setminus\{0\}}\omega^{\tr(ab')}=q\delta_{a,0}-1.
    \end{alignat*}
  \item $i=2$: $\Psi_{(a',b')}$ with $a'\ne 0$ is a non-trivial
    character of $\F_q\times\F_q$. Then
    \begin{alignat*}{5}
      \sum_{x\in P_2}\Psi_x(y)
      &{}=\sum_{a'\in\F_q\setminus\{0\}}\sum_{b'\in\F_q}\Psi_{(a',b')}(y)
     &&{}=\sum_{a'\in\F_q\setminus\{0\}}\sum_{b'\in\F_q}\omega^{\tr(ab'-ba')}\\
      &{}=\sum_{a'\in\F_q\setminus\{0\}}\omega^{\tr(-ba')}\sum_{b'\in\F_q}\omega^{\tr(ab')}
     &&{}=(q\delta_{b,0}-1)q\delta_{a,0}. 
    \end{alignat*}
\end{itemize}
This shows that the sum $\sum_{x\in P_i}\Psi_x(y)$ only depends on the
index $j$ of the partition $P_j$ containing $y$, i.\,e., it is constant on each partition $P_j$.

Using $\Psi_x(y)=\Psi_y(x)^*$, it follows that $\sum_{y\in
  P_j}\Psi_x(y)=\sum_{y\in P_j}\Psi_y(x)^*$, which in turn implies the
second part of the Lemma in \cite{ZiEr09}. Again following
\cite{ZiEr09}, we define
\begin{alignat}{5}
  K_{i,j}=\sum_{x\in P_i}\Psi_x(y),\qquad\text{with $y\in P_j$,}
\end{alignat}
and obtain the matrix
\begin{alignat*}{5}
  K=\begin{pmatrix}
  1 & 1 & 1\\
  q-1 & q-1 & -1\\
  q^2-q & -q & 0
  \end{pmatrix},
\end{alignat*}
which is referred to as Krawtchouk matrix in \cite{GlLu15}.

For the partition of $\F_q\times\F_q$ defined in
\eqref{eq:partition0}--\eqref{eq:partition2}, we set $\mu((a,b))=i$ if
and only if $(a,b)\in P_i$. For a vector $(\bm{a},\bm{b})\in(\F_q\times\F_q)^n$ the
monomial
\begin{alignat*}{5}
  \prod_{i=1}^n Y_{\mu((a_i,b_i))}\in\Z[Y_0,Y_1,Y_2]
\end{alignat*}
encodes how many components of the vector are in $P_0$, $P_1$, and
$P_2$, respectively. For an additive code
$\mathcal{C}\le(\F_q\times\F_q)^n\cong\F_{q^2}^n$, we define the
following \emph{subfield weight enumerator}:
\begin{alignat*}{5}
  \mathcal{W}_{\mathcal{C}}(Y_0,Y_1,Y_2)=\sum_{(\bm{a},\bm{b})\in \mathcal{C}}\prod_{i=1}^n  Y_{\mu((a_i,b_i))}
  =\sum_{i=0}^n\sum_{j=0}^{n-i} A_{i,j} Y_0^{n-i-j}Y_1^i Y_2^j\in\Z[Y_0,Y_1,Y_2].
\end{alignat*}
Using \cite[Theorem 3.5]{GlLu15}, we obtain the following MacWilliams identity.
\begin{thm}
  The subfield weight enumerator
  $\mathcal{W}_{\mathcal{C}^*}(Y_0,Y_1,Y_2)$ of the dual code
  $\mathcal{C}^*$ of $\mathcal{C}$ with respect to the
  trace-symplectic form \eqref{eq:symplectic_form} is obtained from
  the enumerator $\mathcal{W}_{\mathcal{C}}(Y_0,Y_1,Y_2)$ via the MacWilliams
  identity
  \begin{align*}
    \mathcal{W}_{\mathcal{C}^*}(Y_0,Y_1,Y_2)=\frac{1}{|\mathcal{C}|}\mathcal{W}_{\mathcal{C}}(Y_0^*,Y_1^*,Y_2^*),
  \end{align*}
  where
  \begin{align*}
    \left(Y_0^*, Y_1^*, Y_2^*\right)
    &{}=
    \left(Y_0, Y_1, Y_2\right) K
  = \left(Y_0, Y_1, Y_2\right)
    \begin{pmatrix}
      1 & 1 & 1\\
      q-1 & q-1 & -1\\
      q^2-q & -q & 0
    \end{pmatrix}\\
    &{}=\left(
        Y_0+(q-1)Y_1+(q^2-q)Y_2,\;
        Y_0+(q-1)Y_1-q Y_2,\;
        Y_0-Y_1
       \right).
  \end{align*}
\end{thm}

\begin{ex}
  For the symplectic dual $\mathcal{C}^*$ of the MDS code $\mathcal{C}=[17,13,5]_{16}$
  from Example \ref{ex:MDS16}, the subfield weight enumerator is given by
  \begin{small}
  \begin{alignat*}{5}
    {\mathcal W}_{\mathcal{C}^*}=&Y_0^{17}
    && + 816\;Y_0^3Y_1^6Y_2^8 + 3264\;Y_0^3Y_1^4Y_2^{10} + 5304\;Y_0^3Y_1^2Y_2^{12} + 816\;Y_0^3Y_2^{14}\\
    &&&+ 1224\;Y_0^2Y_1^5Y_2^{10} + 1632\;Y_0^2Y_1^3Y_2^{12} + 1224\;Y_0^2Y_1Y_2^{14}\\
    &&&+ 714\;Y_0Y_1^8Y_2^8 + 2448\;Y_0Y_1^6Y_2^{10} + 13056\;Y_0Y_1^4Y_2^{12}\\
    &&&\quad+ 13056\;Y_0Y_1^2Y_2^{14} + 1581\;Y_0Y_2^{16}\\
    &&&+ 1224Y_1^7Y_2^{10} + 5712Y_1^5Y_2^{12} + 9384Y_1^3Y_2^{14} + 4080Y_1Y_2^{16}.
  \end{alignat*}
  \end{small}%
  Using the MacWilliams identity, we obtain the symmetrised partition
  weight enumerator of the code $\mathcal{C}$, given by
  \begin{small}
  \begin{alignat*}{5}
    {\mathcal W}_{\mathcal{C}}&=Y_0^{17}
     &&+ 612\;Y_0^{12}Y_1^4Y_2 + 4488\;Y_0^{12}Y_1^3Y_2^2 + 19992\;Y_0^{12}Y_1^2Y_2^3\\
     &&&\quad+ 36924\;Y_0^{12}Y_1Y_2^4 + 30804\;Y_0^{12}Y_2^5\\
     &&&+ 3264\;Y_0^{11}Y_1^5Y_2 + 33864\;Y_0^{11}Y_1^4Y_2^2 + 163200\;Y_0^{11}Y_1^3Y_2^3\\
     &&&\quad + 499800\;Y_0^{11}Y_1^2Y_2^4 + 809472\;Y_0^{11}Y_1Y_2^5 + 532440\;Y_0^{11}Y_2^6\\
     &&&+ 1224\;Y_0^{10}Y_1^7 + 18360\;Y_0^{10}Y_1^6Y_2 + 212568\;Y_0^{10}Y_1^5Y_2^2\\
     &&&\quad+ 1474512\;Y_0^{10}Y_1^4Y_2^3 + 5842560\;Y_0^{10}Y_1^3Y_2^4\\
     &&&\quad + 14081304\;Y_0^{10}Y_1^2Y_2^5 + 18701496\;Y_0^{10}Y_1Y_2^6 + 10718976\;Y_0^{10}Y_2^7+\ldots
  \end{alignat*}
  \end{small}%
  Here we have omitted terms that correspond to codewords of Hamming
  weight larger than $7$. The subfield weight enumerator
  allows us to deduce the minimal BR-weights given in Example \ref{ex:MDS16}.

\end{ex}

\begin{ex}
For the symplectic dual $\mathcal{C}^*$ of the cyclic code $\mathcal{C}=[17,9,7]_4$ from Example \ref{ex:GF2_17}, the subfield weight enumerator is given by
\begin{small}
\begin{alignat*}{5}
\mathcal{W}_{\mathcal{C}^*}&=
Y_0^{17} &&+ 68Y_0^9Y_1^6Y_2^2 + 476Y_0^9Y_1^4Y_2^4 + 884Y_0^9Y_1^2Y_2^6 + 102Y_0^9Y_2^8\\
&&&+ 1088Y_0^7Y_1^6Y_2^4 + 3536Y_0^7Y_1^4Y_2^6 + 3264Y_0^7Y_1^2Y_2^8 + 272Y_0^7Y_2^{10}\\
&&&+ 68Y_0^5Y_1^{10}Y_2^2 + 680Y_0^5Y_1^8Y_2^4 + 5712Y_0^5Y_1^6Y_2^6 + 12376Y_0^5Y_1^4Y_2^8\\
&&&\quad+ 6460Y_0^5Y_1^2Y_2^{10} + 408Y_0^5Y_2^{12}\\
&&&+ 136Y_0^3Y_1^{10}Y_2^4 + 2040Y_0^3Y_1^8Y_2^6 + 7888Y_0^3Y_1^6Y_2^8 + 10336Y_0^3Y_1^4Y_2^{10}\\
&&&\quad+ 3944Y_0^3Y_1^2Y_2^{12} + 136Y_0^3Y_2^{14}\\
&&&+ 204Y_0Y_1^{10}Y_2^6 + 680Y_0Y_1^8Y_2^8 + 2380Y_0Y_1^6Y_2^{10} + 1836Y_0Y_1^4Y_2^{12}\\
&&&\quad+ 544Y_0Y_1^2Y_2^{14} + 17Y_0Y_2^{16}.
\end{alignat*}
\end{small}%
Using the MacWilliams identity, we obtain the symmetrised partition weight enumerator of the code $\mathcal{C}$, given by
\begin{small}
\begin{alignat*}{5}
\mathcal{W}_{\mathcal{C}}&=
Y_0^{17}&&+ 68Y_0^{10}Y_1^5Y_2^2 + 204Y_0^{10}Y_1^4Y_2^3 + 136Y_0^{10}Y_1^3Y_2^4 + 544Y_0^{10}Y_1^2Y_2^5 \\
&&&\quad+204Y_0^{10}Y_1Y_2^6 + 68Y_0^{10}Y_2^7\\
&&&+ 68Y_0^9Y_1^6Y_2^2 + 476Y_0^9Y_1^4Y_2^4 + 884Y_0^9Y_1^2Y_2^6 + 102Y_0^9Y_2^8+\ldots
\end{alignat*}
\end{small}%
Here we have omitted terms that correspond to codewords of Hamming weight larger than $8$. The subfield weight enumerators allows us to deduce the minimal BR-weights given in Example \ref{ex:GF2_17}.
\end{ex}

We close this section with a remark on the general case of codes over $\F_{q^m}$, for $m>2$.  Using the same approach as above, for codes that distinguish between errors in $\F_q$ and errors in $\F_{q^m}\setminus \F_q$, the resulting partition of the alphabet is no-longer self-dual. More precisely, the subfield weight corresponds to a partition of $\F_{q^m}$ considered as an $\F_q$-vector space into $V_0=\{0\}$, $V_1\setminus V_0$, and $\F_{q^m}\setminus V_1$, where $V_1$ is the one-dimensional space $\F_q$. For the dual partition, we have the chain of vector spaces $V_0< V_{m-1} < V_m=\F_{q^m}$, where $\dim V_{m-1}=m-1$. Hence, for $m>2$, we have to use different weight functions for the code and its dual.

\section{Conclusion}\label{sec:concl}

We introduced two new weights on $\F_{q^m}$, the $\lambda$-subfield weight, and the base-roof weight, both distinguishing between non-zero elements from the base field $\F_q$, and elements outside of it. The former gives rise to a metric, called the $\lambda$-subfield metric. 
If the parameter $\lambda$ is not equal to one, this allows for asymmetric error correction. In particular, if $\lambda$ is greater than one, one can correct more errors in the base field than errors outside of the base field. This particular asymmetric weighting of the domain of the entries is new and can be useful for the construction of quantum codes, where we distinguish diagonal and non-diagonal errors
that occur with different probabilities. 

We gave a theoretical framework for the weights, showing that the $\lambda$-subfield metric is indeed a metric on $\F_{q^m}^n$, deriving upper and lower bounds on the cardinality of codes with a prescribed minimum $\lambda$-subfield distance, and giving an example of optimal codes for this metric (through showing that optimal codes in the rank metric are also optimal in the $\lambda$-subfield metric). Furthermore, we derived a MacWilliams-type identity for the weight enumerator in the case of quadratic field extensions. 

In future work we would like to use this metric for the construction of applicable quantum codes, exploiting the asymmetric error correction capability, which should lead to more efficient quantum error correction than currently known.

\section*{Acknowledgements}

The authors would like to thank Joachim Rosenthal for co-organising the Oberwolfach Workshop 1912 `Contemporary Coding Theory' where the authors first met. Furthermore, they would like to thank the anonymous reviewers for their constructive comments on the paper, in particular for their inspiration for the proof of the non-linear Plotkin bound.

The `International Centre for Theory of Quantum Technologies' project
(contract no. MAB/2018/5) is carried out within the International
Research Agendas Programme of the Foundation for Polish Science
co-financed by the European Union from the funds of the Smart Growth
Operational Programme, axis IV: Increasing the research potential
(Measure 4.3). 

The third author  is  supported by the European Union's Horizon 2020 research and innovation programme under the Marie Sk\l{}odowska-Curie grant agreement no. 899987.

\bibliographystyle{plain}
\bibliography{main}

\end{document}